%Paper: cond-mat/9509048
%From: zhongd@craft.camp.clarkson.edu (Dexin Zhong)
%Date: Sat, 9 Sep 1995 18:02:58 -0400

\magnification=1200
\overfullrule = 0 pt
\baselineskip=24 true bp
\hfuzz=0.5pt

\def\etal{{\it et al.\/}}
\def\ie{{\it i.e.\/}}

\def\av#1{{\langle{#1}\rangle}}
\def\NI{\noindent}

\centerline{\bf Universality Class of Two-Offspring}
\centerline{\bf Branching Annihilating Random Walks}

\vskip 1cm
\centerline{Dexin Zhong\footnote{$^1$}{e-mail:
{\sl zhongd@craft.camp.clarkson.edu}}}
\medskip
\centerline{Physics Department, Clarkson University,
Potsdam, New York 13699--5820, USA}
\bigskip
\centerline{Daniel ben-Avraham\footnote{$^2$}{e-mail:
{\sl qd00@craft.camp.clarkson.edu}}}
\medskip
\centerline{Physics Department, Clarkson University,
Potsdam, New York 13699--5820, USA}
\centerline{and}
\centerline{Physics Department, Bar-Ilan University, Ramat-Gan 52900, Israel}
\vskip 1.2cm
\NI{\bf ABSTRACT:} We analyze a two-offspring
Branching Annihilating Random Walk ($n=2$ BAW) model,
with finite annihilation rate.
The finite annihilation rate allows for a dynamical phase transition between
a vacuum, absorbing state and a non-empty, active steady state.
We find numerically
that this
transition belongs to the same universality class as BAW's with an even number
of offspring, $n\geq 4$, and that of other
models whose dynamic rules conserve the parity of the particles locally.
The simplicity of the model is exploited in computer simulations to obtain
various critical exponents with a high level of accuracy.

\bigskip
\NI PACS:\ \ 64.60.--i\ \ 02.50.--r\ \ 05.70.Ln,\ \ 82.20.Mj
\vfill\eject

% INTRODUCTION
%
Nonequilibrium phase transitions in dynamical lattice models are the subject
of rapidly growing interest [1,2].  Most second-order (\ie, continuous)
phase transitions studied to date seem to belong to the same universality
class as directed percolation (DP) [3-14].
An early observation of this fact
has led Grassberger [15] and Janssen [16] to the conjecture that all
continuous phase transitions from a single absorbing state to an active steady
state seen in one-component models are in the DP universality class.

The DP conjecture draws support from the field-theoretical
description of many of the models in question.  In a coarse-grained fashion,
this turns out to be identical to Reggeon Field Theory (RFT) [17].
An important achievement of Cardy and Sugar was proving
that DP and RFT are in the same universality class [18].

The DP universality class is extremely robust.  Examples not only
include
one-component models with a transition about a single absorbing state (as
postulated in the conjecture) with binary interactions [3-6] and higher-order
interactions [5], but also models with many absorbing states [7] and
multi-component systems [8-11].  Grinstein \etal\ [8] have presented
theoretical arguments
for the extension of the DP conjecture to the latter case.
In fact, exceptions to DP transitions are rare.  In several cases, models
that were initially thought to violate the rule were later
found to be in the DP class, upon more careful analysis [12-14].

Recently, the study of Branching Annihilating Walks (BAW's) has revealed
a new universality class.
The BAW model consists of particles which diffuse on a lattice and
annihilate immediately upon encounter.  Each particle gives
birth to $n$ other offspring at adjacent sites, at some prescribed rate.
The situation may be summarized, schematically, by
$$
\eqalign{
A+A &\to 0, \cr
A   &\to (1+n)A. \cr } \eqno(1)
$$
As the birth (or branching) rate increases, the observed long time behavior
undergoes a phase transition, from a state where the system is empty to an
active steady state with a finite concentration of particles.
For odd $n$, the transition is in the same class as DP [13], but a
new class emerges for even $n\geq 4$ [19,20].  Because the number of particles
modulo 2 is conserved when $n$ is even, this class is referred
to as the ``parity-conserving universality class'' (PC).
Other systems in the PC class include Grassberger's cellular automata models
A [21] and B [22],
a kinetic Ising model of Menyh\'ard [23], and the more complex (two-component)
interacting monomer-dimer model of Kim and Park [24].

In the original BAW model the annihilation rate is infinite.  In this case,
the $n=2$ BAW does not undergo a transition.  It always evolves into the
vacuum absorbing state, regardless of the branching rate [25].
However, a transition may be observed upon the introduction of a finite
reaction rate [23,26].  Below, we
report results for the phase transition of the $n=2$ BAW with finite
reaction rate.  We find that it is in the PC class, and the simplicity of
the model ($n=2$ offspring, as opposed to even $n\geq 4$) allows us to
obtain accurate estimates of various critical exponents.

% THE MODEL
%
We define our model on a one dimensional lattice by the following computer
algorithm.  A particle is picked at random.  It can undergo either
nearest-neighbor diffusion (with probability $1-p$) or it may branch (with
probability $p$).
In a diffusion attempt, a random direction (left or right) is picked and the
particle moves to the target site if it is unoccupied.  If the target site
is occupied, then annihilation of the incident and target particles occurs with
probability $k$.  Otherwise, the incident particle remains at its original
position.  If a branching attempt is selected, then branching
to the two nearest-neighbor sites occurs with probability 1 if both neighbors
are empty.  If one or both neighbors are occupied, branching to both
neighbors occurs with probability $k$.  In this case, if a new
particle is placed on a previously occupied site, then both particles are
removed.  The probability $k$ controls the annihilation rate ($k=1$ corresponds
to the original model of infinte reaction rate).

Each time a particle is picked, a ``time" counter is incremented by $1/N$,
where $N$ is the number of currently surviving particles.  Thus, in a time step
each particle either branches or diffuses one time, on the average.
The ratio of branching rate to diffusion rate is controlled by the parameter
$p$.  Both $p$ and $k$ are critical fields:
a phase transition could be observed upon varying either of the two.
We chose to fix the reaction probability at $k=1/2$, and to study the
dynamical phase transition as a function of $p$.

% RESULTS
%
To determine the transition point, we follow the common technique of
epidemic analysis [27,20].  Starting our simulations at $t=0$ with an empty
lattice, except two adjacent occupied sites placed at the origin, we let
the system evolve.  The two-particle seed may grow and spread, but the
system may also hit the absorbing state, when all particles annihilate.
We measure: (1) $P(t)$, the probability that the system has not entered the
absorbing state up to time $t$, (2) $\av{n(t)}$, the mean number of
particles at time $t$, averaged over all runs,
and (3) $\av{R^2(t)}$, the mean square distance
of spreading from the center of the lattice, averaged over only the surviving
runs.  When the branching rate is critical, $p\to p_c$, these quantities
are governed by power-laws in the long time asymptotic limit:
$$
\eqalign{
P(t)        & \sim t^{-\delta}, \cr
\av{n(t)}   & \sim t^{\eta}, \cr
\av{R^2(t)} & \sim t^z. \cr     }  \eqno(2)
$$
Off criticality, these quantities either grow or decay exponentially fast.
As a result, a plot of the local slopes
$$
-\delta(t)={\ln[P(t)/P(t/m)]\over\ln m} \eqno(3)
$$
(and similarly for $\eta$ and $z$) as a function of $1/t$ is linear at
criticality---and extrapolates to the long time asymptotic value---but shows
a telltale curvature, otherwise.  In Eq.~(3), $m>1$ is an arbitrary factor
which we took to be equal to 5.

Running concurrently on a cluster of about 30 RISC/6000 workstations, we
were able to perform extensive computer simulations.  For each value of $p$,
about $10^8$ runs were performed, up to 50,000 time steps each.  The size
of the lattice was chosen such that the spreading seed never hit the
boundaries, to avoid finite-size effects.

In Fig.~(1), we plot the local slope $\eta(t)$
for various values of $p$.  One clearly sees that the curves for $p=0.4960$
and  $p=0.04950$ veer upward, and the curves for $p=0.4942$ and $p=0.4920$
veer downward, while the curve for $p=0.4946$ shows no significant curvature.
This, combined with the fact that for the PC universality class $\eta$
is expected to be zero [20], helps us determine the critical value
at $p_c=0.4946(2)$.
Without relying on the assumption that $\eta=0$, we estimate from our data that
$\eta=0.000(1)$.

In Figs.~(2) and (3), we show analogous results for $\delta(t)$ and $z(t)$.
The same trend in the curvature of the local slopes is observed, though the
plots for $\eta$ (Fig.~1) allow for a more conclusive determination
of the critical point.  From the data we conclude that $\delta=0.286(2)$
and $z=1.147(4)$.

A critical exponent of particular interest is $\beta$, the order-parameter
critical exponent.  This is related to the stationary concentration of
particles in the steady state, $\rho$, through
$$
\rho \sim (p-p_c)^{\beta}, \qquad (p_c<p).  \eqno(4)
$$
To measure $\beta$, we start a simulation with a fully occupied lattice and
let the system evolve until it reaches a steady state.  We then average
the concentration over a long period of time.  In this fashion, we have
measured $\rho$ for various values of $p$.  Each data point was obtained from
over 1000 independent runs on $10^4$-site lattices, averaged
over periods of up to $t=10^7$ time steps.  In Fig.~(4), we show a log-log
plot of the steady state concentration as a function of $p-p_c$.  From
the slope of the data we estimate $\beta=0.922(5)$.

In Table~1, we compare our critical exponents measured for the $n=2$ BAW
with finite reaction rate to the exponents measured for the $n=4$ BAW
with infinite reaction rate [20].  (We anticipate that making the reaction
rate finite wont change the nature of the transition in the latter case.)
There is excellent agreement between the two sets of measurements, suggesting
that our model is in the PC universality class.  Similar figures, but
with larger numerical uncertainties, have been obtained for the other
models in the PC class [21-24].  The relative simplicity of the present model
allows for efficient data
collection.  Indeed, we have achieved a modest improvement in the estimate
of $\beta$, the order-parameter critical exponent.

% DISCUSSION
%
The new PC universality class remains poorly understood.  For example,
one-component models in this class (such as BAW's
and Grassberger's automata) may be treated through a second-quantization
formalism, by means of creation-annihilation operators [28].  The
equations thus derived may be coarse-grained through standard renormalization
group techniques.  This results in RFT equations,
indicating that these systems should be in the DP universality class.
What happens is that crucial parity-conserving fields renormalize away, and do
not show in the final analysis.

Is parity conservation solely responsible for the new PC universality class?
Undeniably, it is a feature of all the models found to be in the PC
class so far, and its neglect in the field-theoretical approach leads to
erroneous conclusions.  Nevertheless, Park and Park [29] have convincingly
proved, at least for the case of the interacting monomer-dimer model,
that it is the existence of equivalent groundstates (in a dynamical sense,
\ie, states that the system may access with equal probability), and not
parity conservation, that gives rise to the new PC class.  It would be
interesting to see if this can be shown for other models in the PC class.

The BAW model analyzed in this paper is dual to the non-equilibrium Ising
kinetic model of Menyh\'ard.  The particles in our model correspond to
the ``kinks", or the boundaries between domains of oppositely oriented spins.
With only one component, and local dynamic rules involving a small number
of particles, the $n=2$ BAW is one of the simplest models in the PC
universality class.
We hope that this will be exploited in future numerical work, to obtain
better estimates of the critical exponents.  An intriguing possibility
raised by Jensen [20], is that the PC transition may be solved analytically.
It would make sense to search for analytical solutions starting
with the simplest models, such as the $n=2$ BAW.

% ACKNOWLEDGMENTS
%
We thank Hyunggyu Park for numerous useful discussions.

\centerline{\bf References}
\medskip

\item{1)} G. Nicolis and I. Prigogine, {\sl Self-organization in
Nonequilibrium Systems} (Wiley Interscience, New York, 1977).
\smallskip
\item{2)} H. Haken, {\sl Synergetics} (Springer-Verlag, New York, 1983).
\smallskip
\item{3)} T. E. Harris, {\sl Ann. Prob.} {\bf 2}, 969 (1974).
\smallskip
\item{4)} R. Dickman and M. A. Burschka, {\sl Phys. Lett. A} {\bf 127},
132 (1988).
\smallskip
\item{5)} R. Dickman, {\sl Phys. Rev. B} {\bf 40}, 7005 (1989).
\smallskip
\item{6)} D. A. Browne and P. Kleban, {\sl Phys. Rev. A} {\bf 40},
1615 (1989); T. Aukrust, D. A. Browne, and I. Webman,
{\sl Europhys. Lett.} {\bf 10}, 249 (1989); {\sl Phys. Rev. A} {\bf 41},
5294 (1990).
\smallskip
\item{7)} I. Jensen and R. Dickman, {\sl Phys. Rev. E} {\bf 48}, 1710 (1993).
\smallskip
\item{8)} G. Grinstein, Z.-W. Lai, and D. A. Browne,
{\sl Phys. Rev. A} {\bf 40}, 4820 (1989).
\smallskip
\item{9)} R. M. Ziff, E. Gulari, and Y. Barshad, {\sl Phys. Rev. Lett.}
{\bf 56}, 2553 (1986).
\smallskip
\item{10)} I. Jensen, H. C. Fogedby, and R. Dickman, {\sl Phys. Rev. A}
{\bf 41}, 3411 (1990).
\smallskip
\item{11)} H. Park, J. K\"ohler, I.-M. Kim, D. ben-Avraham, and S. Redner,
{\sl J. Phys. A} {\bf 26}, 2071 (1993).
\smallskip
\item{12)} R. Bidaux, N. Boccara, and H. Chat\'e, {\sl Phys. Rev. A} {\bf 39},
3094 (1989); I. Jensen, {\sl Phys. Rev. A} {\bf 43}, 3187 (1991).
\smallskip
\item{13)} H. Takayasu and A. Y. Tretyakov, {\sl Phys. Rev. Lett.} {\bf 68},
3060 (1992); I. Jensen, {\sl Phys. Rev. E} {\bf 47}, 1 (1993).
\smallskip
\item{14)} J. K\"ohler and D. ben-Avraham,
{\sl J. Phys. A} {\bf 24}, L621--L627 (1991);
D. ben-Avraham and J. K\"ohler, {\sl J. Stat. Phys.} {\bf 65}, 839--848 (1991);
I. Jensen, {\sl J. Phys. A} {\bf 27}, L61 (1994).
\smallskip
\item{15)} P. Grassberger, {\sl Z. Phys. B} {\bf 47}, 365 (1982).
\smallskip
\item{16)} H. K. Janssen, {\sl Z. Phys. B} {\bf 42}, 151 (1981).
\smallskip
\item{17)} V. N. Gribov, {\sl Sov. Phys. JETP} {\bf 26}, 414 (1968);
H. D. I. Abarbanel, J. B. Bronzan, R. L. Sugar, and A. R. White,
{\sl Phys. Rep.} {\bf 21C}, 120 (1975); R. C. Brower, M. A. Furman, and
M. Moshe, {\sl Phys. Lett.} {\bf 76B}, 213 (1978).
\smallskip
\item{18)} J. L. Cardy and R. L. Sugar, {\sl J. Phys. A},{\bf 13}, L423 (1980).
\smallskip
\item{19)} I. Jensen, {\sl J. Phys. A} {\bf 26}, 3921 (1993).
\smallskip
\item{20)} I. Jensen, {\sl Phys. Rev. E} {\bf 50}, 3623 (1994).
\smallskip
\item{21)} P. Grassberger, F. Krause, and T. van der Twer, {\sl J. Phys. A}
{\bf 17}, L105 (1984).
\smallskip
\item{22)} P. Grassberger, {\sl J. Phys. A} {\bf 22}, L1103 (1989).
\smallskip
\item{23)} N. Menyh\'ard, {\sl J. Phys. A} {\bf 27}, 6139 (1994).
\smallskip
\item{24)} M. H. Kim and H. Park, {\sl Phys. Rev. Lett.} {\bf 73}, 2579 (1994);
H.~Park, M.~H.~Kim, and H.~Park, (preprint, to appear in {\sl Phys. Rev. E}).
\smallskip
\item{25)} A. Sudbury, {\sl Ann. Prob.} {\bf 18}, 581 (1990)
\smallskip
\item{26)} D. ben-Avraham, F. Leyvraz, and S. Redner,
{\sl Phys. Rev. E} {\bf 50}, 1843 (1994).
\smallskip
\item{27)} P. Grassberger and A. de la Torre, {\sl Ann. Phys. (NY)} {\bf 122},
373 (1979).
\smallskip
\item{28)} M. Doi, {\sl J. Phys. A} {\bf 9}, 1465 and 1479 (1976);
P.~Grassberger and M.~Scheunert, {\sl Fortsch. Phys.} {\bf 28}, 547 (1980);
L.~Peliti, {\sl J. Phys. (Paris)} {\bf 46}, 1469 (1985);
R.~Dickman, {\sl J. Stat. Phys.} {\bf 55}, 997 (1989); I.~Jensen and
R.~Dickman, {\sl Physica A} {\bf 203}, 175 (1994).
\smallskip
\item{29)} H. Park and H. Park, {\sl Critical behavior of an absorbing
phase transition in an interacting monomer-dimer model}, (preprint, to appear
in {\sl Physica A}).

\vfill\eject
\noindent
\centerline{\bf CAPTIONS}
\bigskip
\bigskip
\bigskip

\NI\hang{\bf Figure 1:} Local slope $\eta(t)$ as a function of $1/t$.
Shown are curves for $p=0.4960$, $0.4950$, $0.4946$,
$0.4942$, and $0.4920$ (top to bottom).
\bigskip

\NI\hang{\bf Figure 2:} Local slope $\delta(t)$ as a function of $1/t$.
Shown are curves for $p=0.4950$, $0.4946$,
and $0.4942$ (top to bottom).
\bigskip

\NI\hang{\bf Figure 3:} Local slope $z(t)$ as a function of $1/t$.
Shown are curves for $p=0.4950$, $0.4946$,
and $0.4942$ (top to bottom).
\bigskip

\NI\hang{\bf Figure 4:} Steady state concentration, $\rho$, as a function
of the branching probability, $p-p_c$.  The slope of the data points yields
the order-parameter critical exponent, $\beta$.

\vfill\eject
\centerline{\bf Table 1}
\vskip 0.5in

\settabs 5 \columns
\+       & {$\delta$} & {$\eta$}  &   z       & {$\beta$}     &\cr

\+ $n=4$ & $0.285(2)$ &$0.000(1)$ &$1.141(2)$ &$0.92(3)$      &\cr

\+ $n=2$ & $0.286(2)$ &$0.000(1)$ &$1.147(4)$ &$0.922(5)$     &\cr

\bigskip
\NI\hang{\bf Table 1:} Critical exponents for the $n=4$ BAW
(from Ref. 20)
and for the $n=2$ BAW (as measured in the present work).

\vfill\eject

%%%%%%%%%%%%%%%%%%%%%%%%%%%%%%%%%%%%%%%%%%%%%%%%%%%%%%%%%%%%%%%%%%%%%%%%%%%%%%

%       Delete till \bye if you don't want figures

\input epsf
\epsfverbosetrue
\epsfxsize 300pt
\epsfysize 200pt
\epsfbox{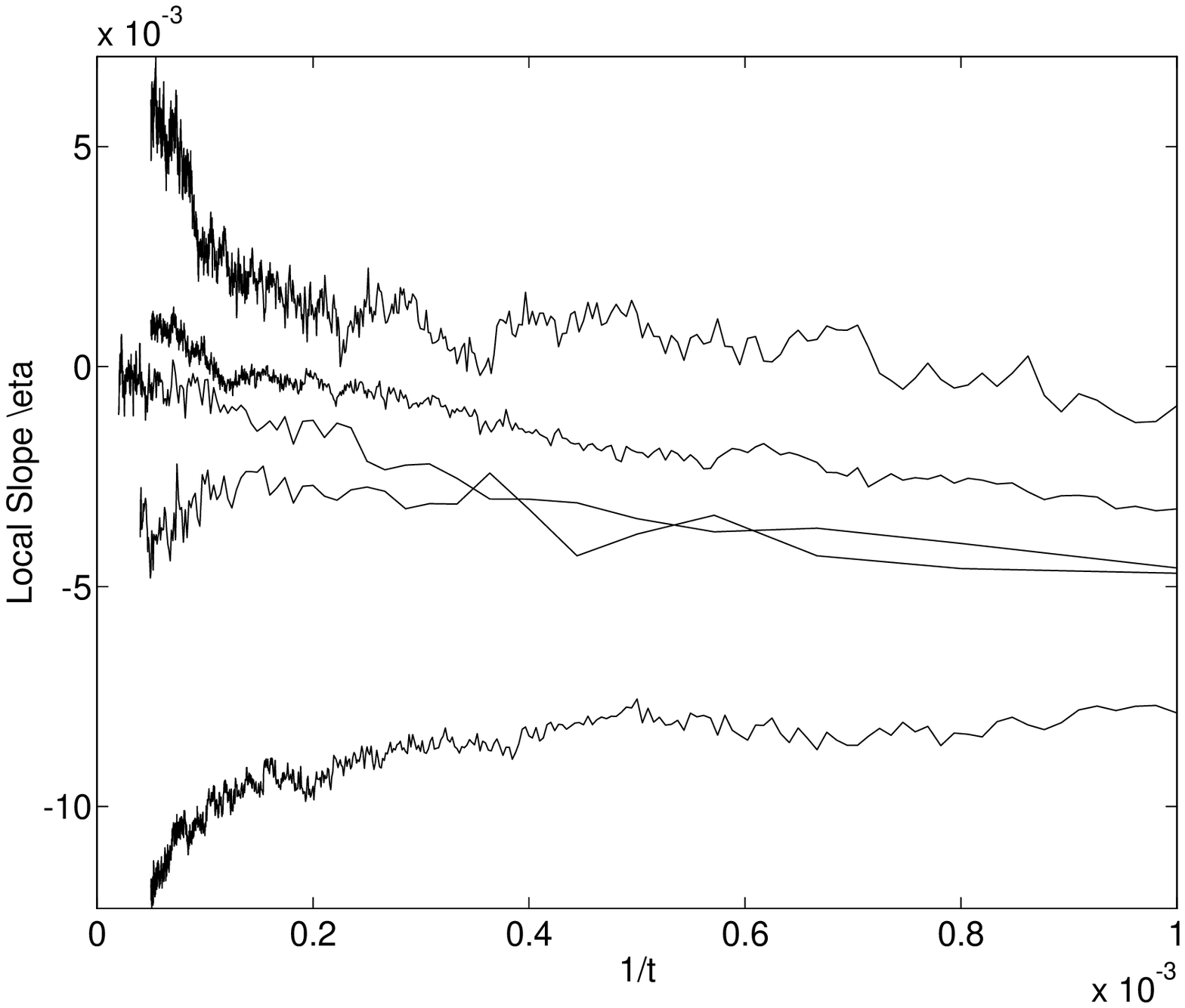}
\epsfxsize 300pt
\epsfysize 200pt
\epsfbox{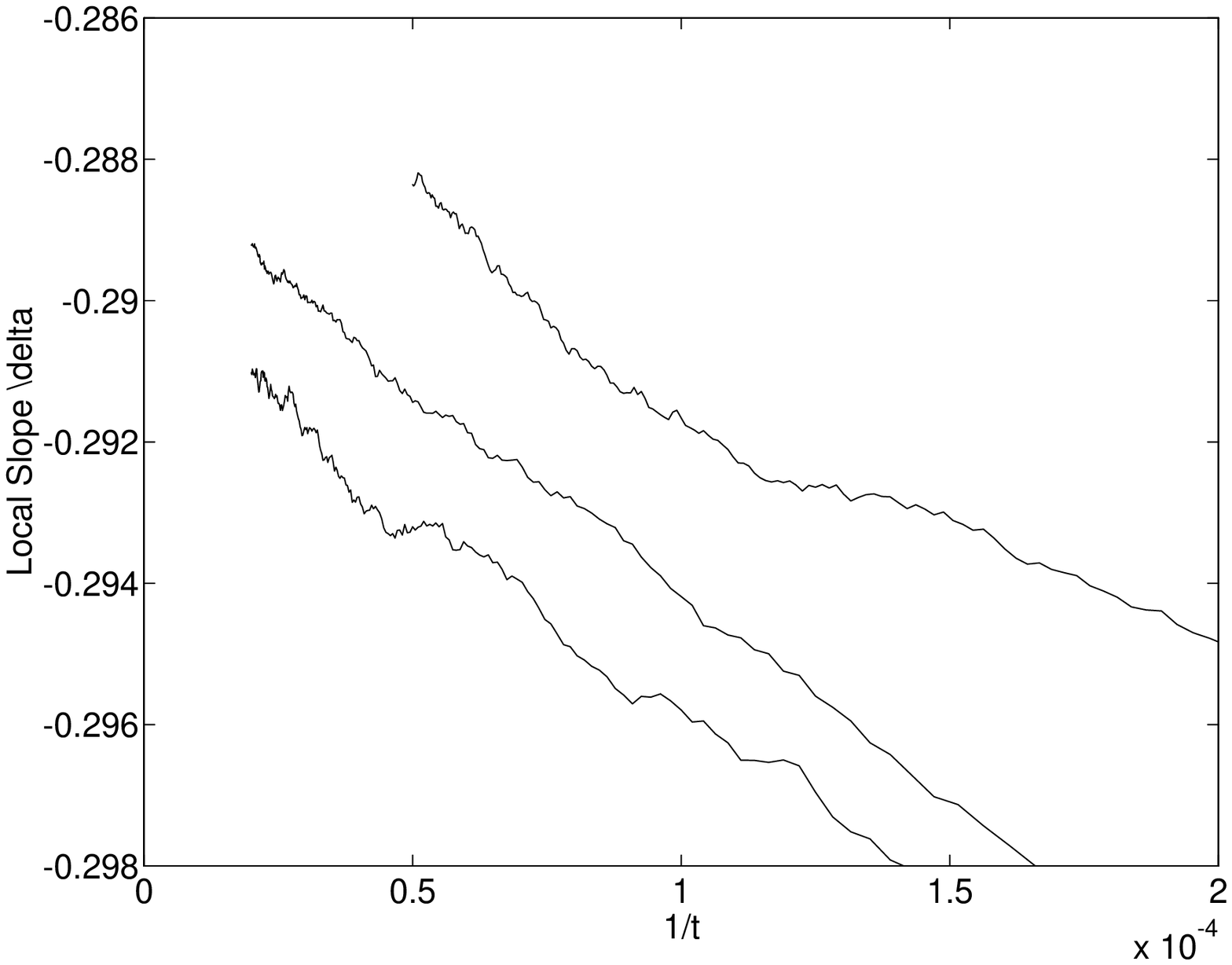}
\epsfxsize 300pt
\epsfysize 200pt
\epsfbox{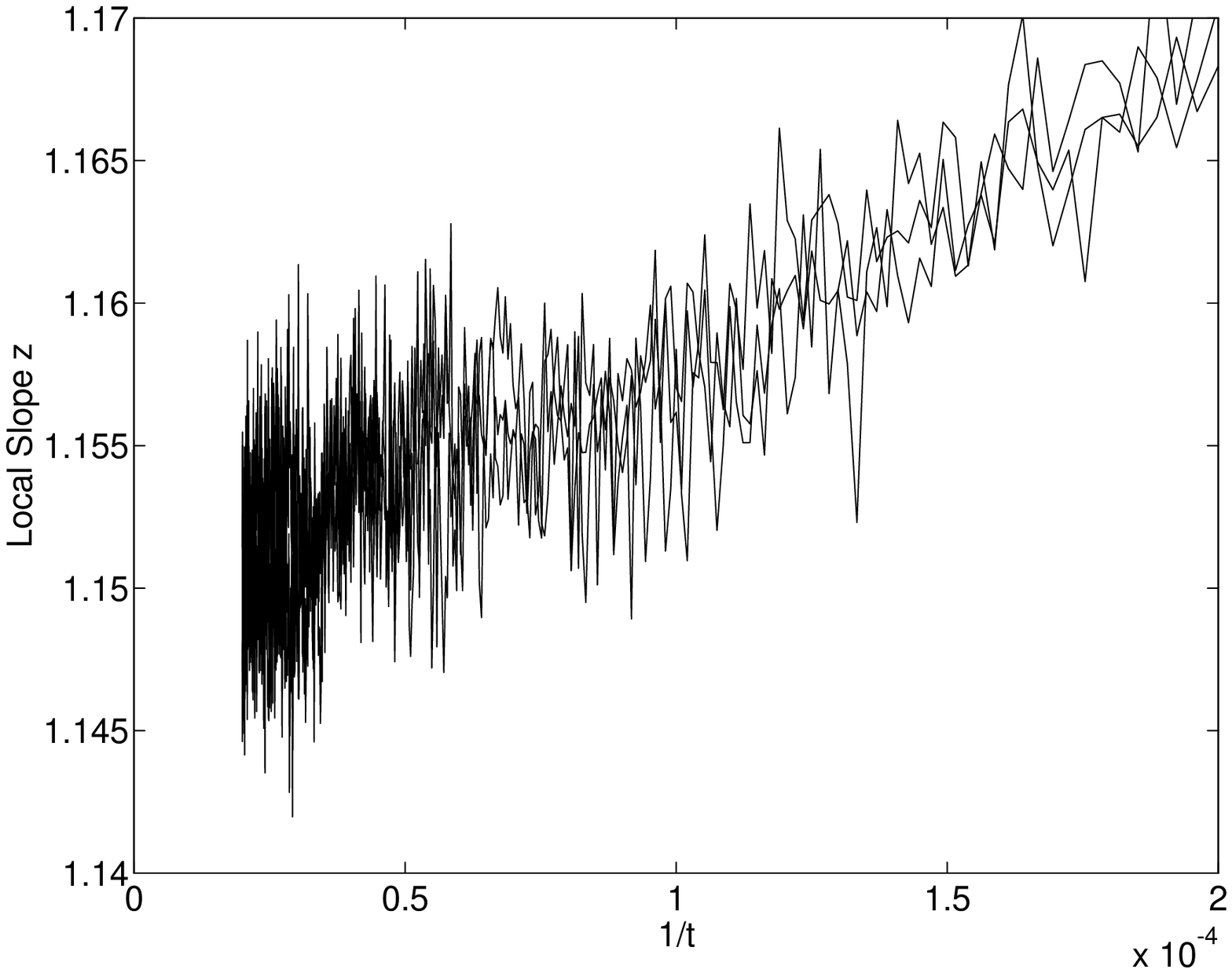}
\epsfxsize 300pt
\epsfysize 200pt
\epsfbox{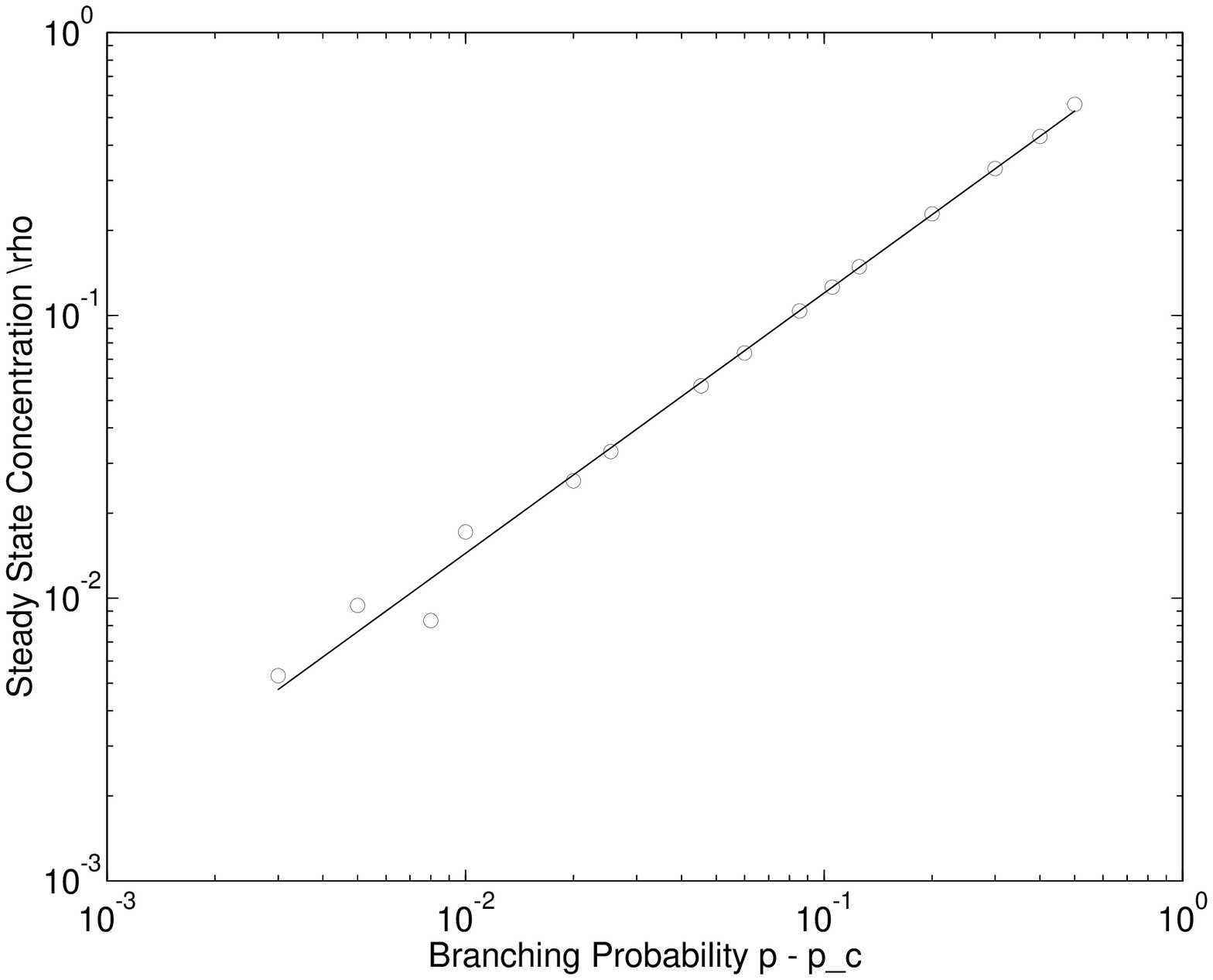}

\bye